\documentclass[a4paper,11pt]{article}
\usepackage{pos}
\usepackage{wrapfig}
\usepackage{siunitx}
\usepackage{hyperref}
\newcommand{\reffig}[1]{Fig.~\ref{#1}}

\newcommand{\refref}[1]{Ref.~\cite{#1}}

\title{Observation of air showers with an IceCube-Gen2 prototype station at the Pierre Auger Observatory}
\ShortTitle{Observation of air showers with an IceCube-Gen2 prototype station at Auger}

\manuallySeparateAuthors
\author*[a]{Stef Verpoest}
\author[1]{for the IceCube-Gen2}
\author{ and }
\author[b,2]{Pierre Auger}
\author{ Collaborations}

\affiliation[a]{Bartol Research Institute, Department of Physics and Astronomy, University of Delaware,\\
  Sharp Lab, 104 The Green, Newark DE, 19716, United States of America}
\affiliation[b]{Observatorio Pierre Auger, Av. San Martín Norte 304, 5613 Malargüe, Argentina}

\note{Full author list at \href{https://icecube.wisc.edu/collaboration/authors/\#collab=IceCube-Gen2\&date=2024-06-13\&formatting=web}{https://icecube.wisc.edu/collaboration/authors}.}
\note{Full author list at \url{http://www.auger.org/archive/authors_2024_06.html}.}

\emailAdd{verpoest@udel.edu}

\abstract{The next-generation neutrino telescope IceCube-Gen2 is planned to include a surface detector array consisting of scintillation detectors and radio antennas for the detection of cosmic-ray air showers. Prototype stations, each comprising 8 scintillator panels and 3 SKALA antennas, have been deployed to various locations. One of these stations is located at the site of the Pierre Auger Observatory, situated within the most densely instrumented part of the surface detector array, which features a spacing of 433 meters.
In this contribution, we present first results from this prototype station, including the observation of the Galactic noise, and the first coincident detection of air showers between the prototype radio antennas and the water-Cherenkov detectors of the Pierre Auger Observatory.}

\FullConference{10th International Workshop on Acoustic and Radio EeV Neutrino Detection Activities (ARENA2024)\\
11-14 June 2024\\
The Kavli Institute for Cosmological Physics, Chicago, IL, USA\\}

\begin{document}
\maketitle

\section{Introduction}

The proposed design for the next-generation neutrino telescope IceCube-Gen2 includes a surface array comprising scintillation detectors and radio antennas for the observation of cosmic-ray air showers~\cite{IceCube-Gen2:2020qha}. A prototype station of surface detectors has been collecting data at the South Pole for several years~\cite{Venugopal}, demonstrating the successful detection of air showers. A second prototype station was deployed at the location of the Pierre Auger Observatory~\cite{PierreAuger:2015eyc} in Argentina and has been taking data continuously since December 2022. It is located in the most densely instrumented part of the surface detector, the SD433 array, which has a spacing of \SI{433}{\m} between the water-Cherenkov detectors~\cite{PierreAuger:2021tmd}. The station is also contained inside the footprint of the Auger Engineering Radio Array (AERA)~\cite{PierreAuger:2018pmw}.

\begin{figure}[b]
    \centering
    \includegraphics[width=0.3\linewidth]{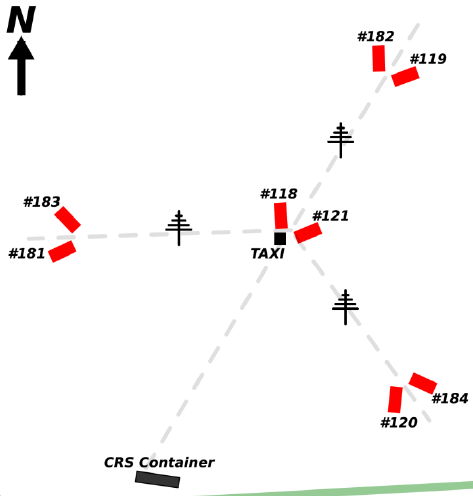}\qquad\includegraphics[trim={10cm 5cm 15cm 10cm},clip, width=0.6\linewidth]{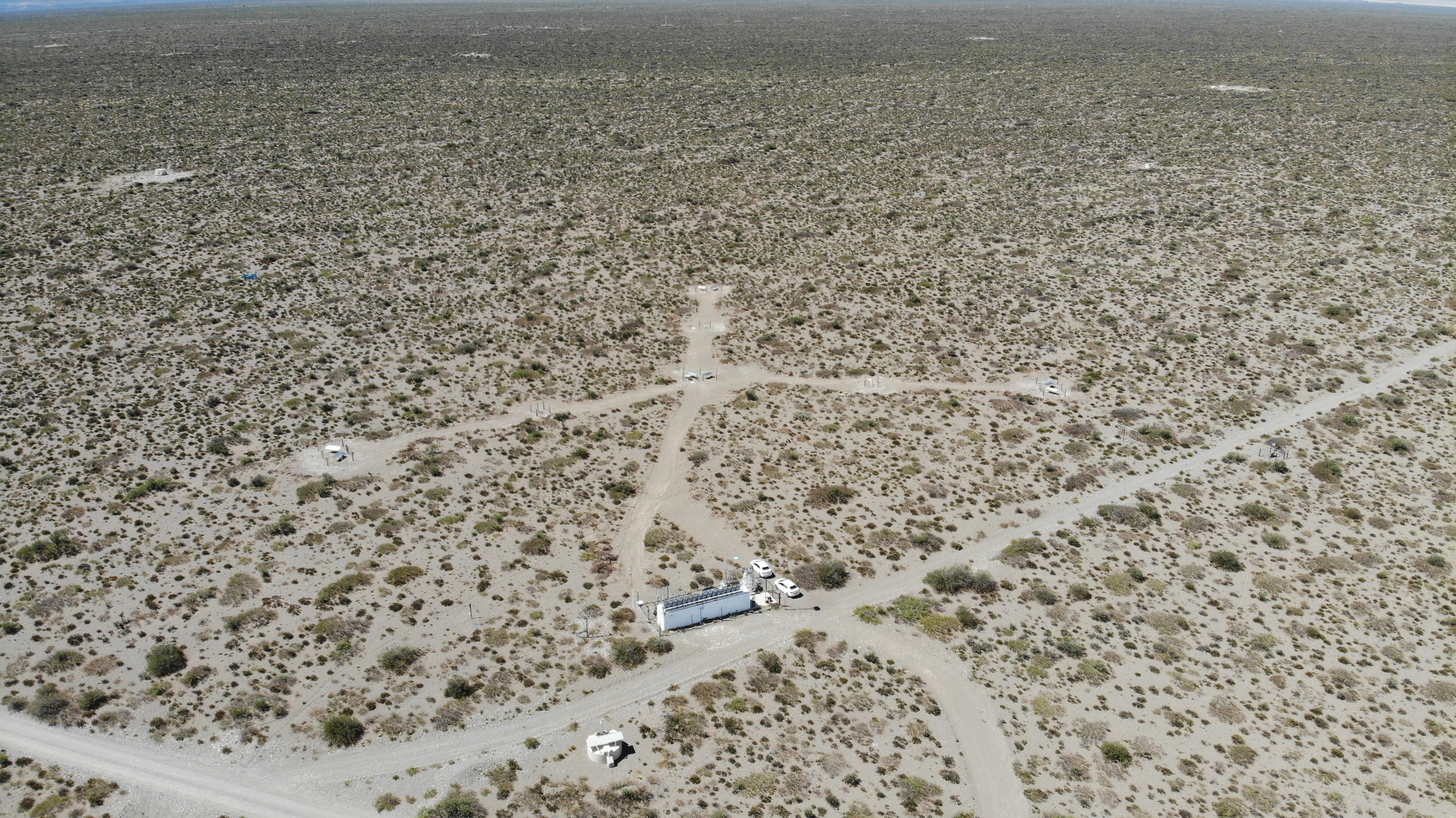}
    \caption{Left: Layout of the prototype detector station. The red rectangles represent scintillator panels, the antennas are shown in black. The central DAQ is represented by the black square labeled ``TAXI''. The Central Radio Station of AERA is used for data collection and power supply. Right: Drone shot of the station at the site of the Pierre Auger Observatory (photo by Tim Huege).}
    \label{fig:picture}
\end{figure}

The layout of the prototype station is shown in \reffig{fig:picture}. Three arms of about \SI{70}{\m} length, each with a pair of scintillator panels at the end, extend outwards from the center, where another pair of scintillators and the data-acquisition system (DAQ) are located. In the middle of each of the arms is a radio antenna of the SKALA-v2 type~\cite{skala}, which has two perpendicular polarizations. For this particular station, the antennas are all oriented differently. The readout of the radio data is triggered by the scintillators. The radio data is also read out at fixed intervals to collect background radio waveforms. Power is provided by the Central Radio Station (CRS) of AERA, where also data collection takes place.

We present here the first results obtained from the data taken with the radio antennas of the prototype station deployed at the Pierre Auger Observatory, including studies of the radio background and the first detection of air showers, confirmed by identifying coincident events in the SD433 data. The radio data were analyzed using the radio-specific modules of IceCube's IceTray framework~\cite{IceCube:2022dcd}.

\section{Radio background measurements}

\begin{figure}[h]
    \centering
    \includegraphics[width=0.535\textwidth]{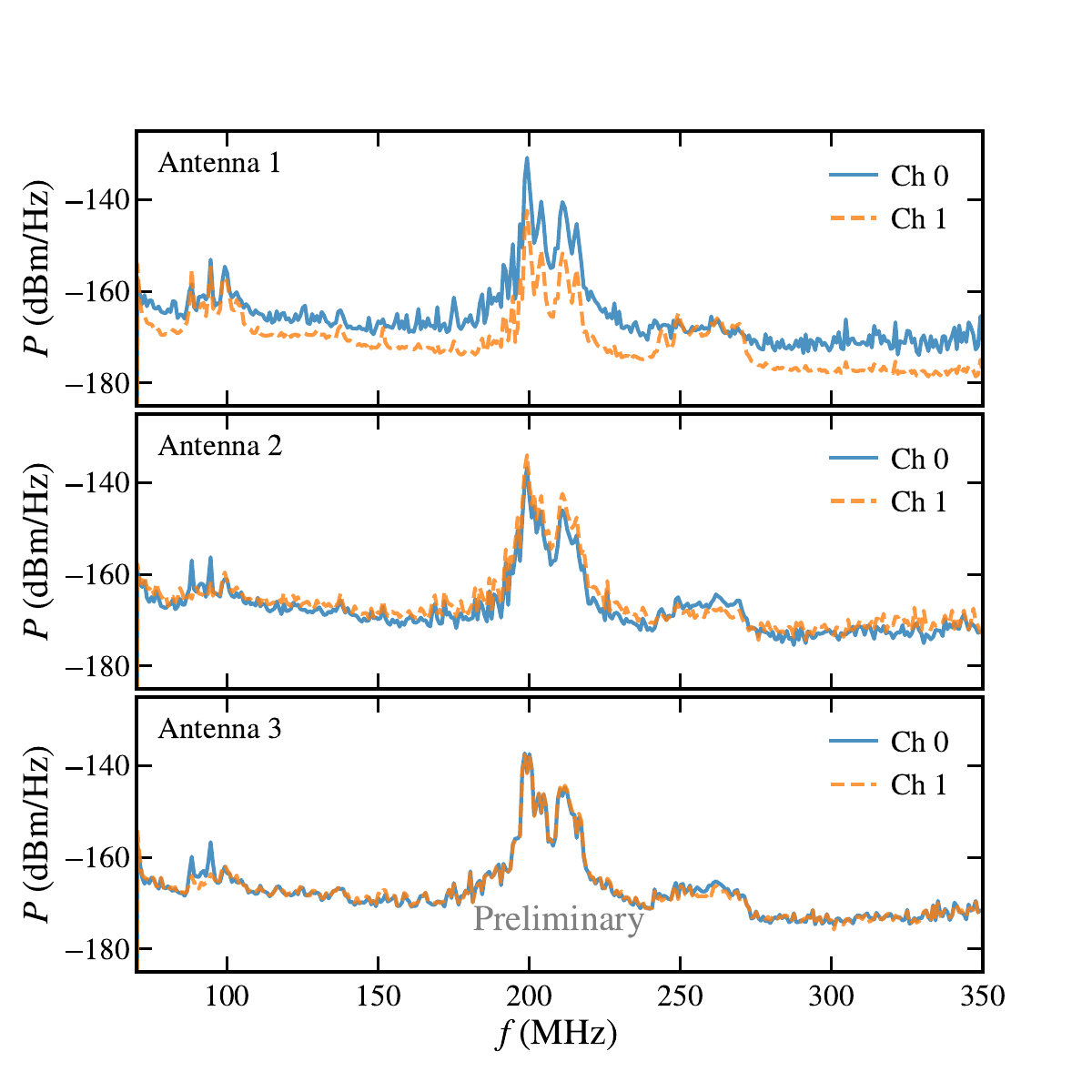}\includegraphics[width=0.465\textwidth]{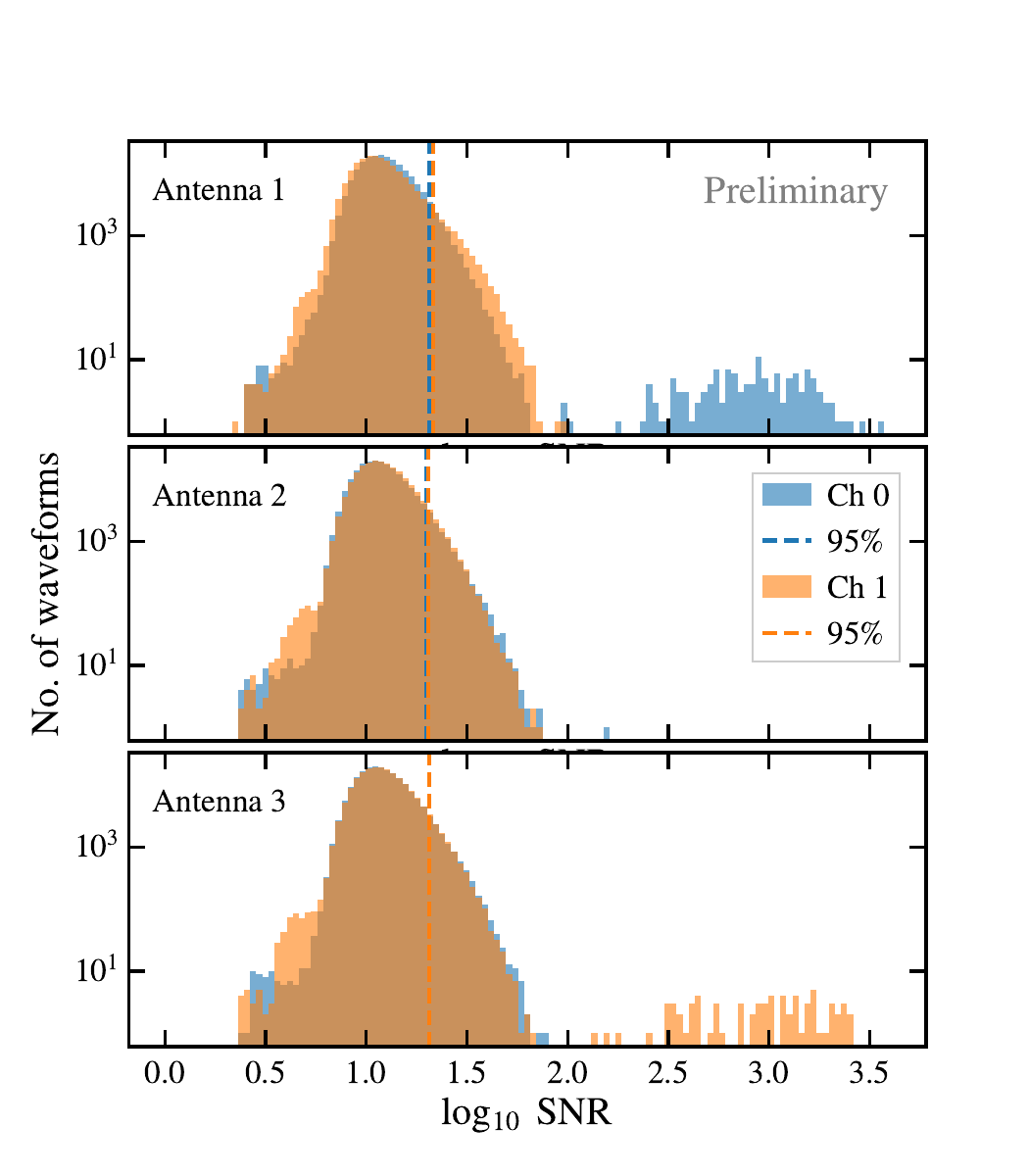}
    \caption{Left: Frequency spectrum of the background traces recorded by the three antennas of the prototype station. Right: Distribution of values of the signal-to-noise ratio calculated for the background waveforms (fixed-rate trigger), after applying a 110-\SI{185}{\mega\Hz} bandpass filter and a filter to suppress RFI. The vertical lines indicate the 95th percentile. The distinct high-SNR population seen in some channels is related to occasional DAQ artifacts.}
    \label{fig:spectrum}
\end{figure}

The radio background is studied using the waveforms obtained with the fixed-rate trigger. The frequency spectrum corresponding to the recorded background waveforms for each polarization of the three antennas, after the removal of the electronics response, is shown in the left panel of \reffig{fig:spectrum}. Large RFI contributions to the spectrum can be seen around \SI{100}{\mega\Hz} and \SI{200}{\mega\Hz}, corresponding to FM radio and TV bands, respectively.

\begin{figure}[t]
    \centering
    \includegraphics[width=\linewidth]{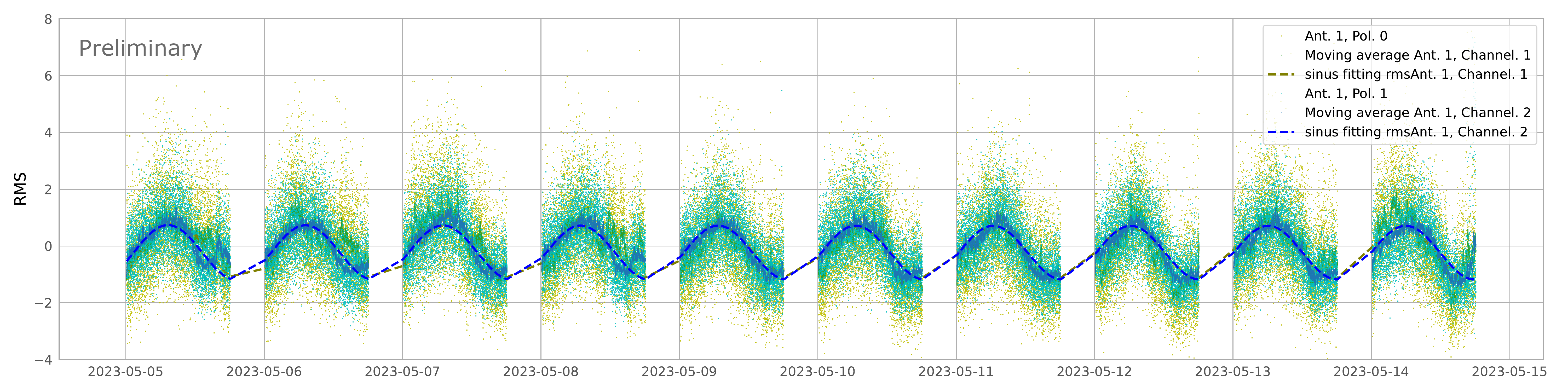}
    \caption{Time dependence of the RMS (and its moving average) of background waveforms. A sinusoidal fit is performed, indicating a predominantly sidereal variation. Figure by Sara Reina.}
    \label{fig:galactic}
\end{figure}

The background waveforms have been used to observe the time variation of the Galactic noise. The root mean square (RMS) was calculated for each waveform after filtering to the relatively clean band between \SI{110}{\mega\Hz} to \SI{130}{\mega\Hz}. The RMS values are shown over 10 days for illustration in \reffig{fig:galactic}, where a periodic behavior can be observed. By fitting a moving average of the RMS over a long period of time, it was confirmed that it shows a predominantly sidereal variation, as expected for background noise related to the movement of the Galaxy across the sky.

\section{Coincident detection of air showers with radio antennas and Auger SD433}

In this section, we describe a search for air showers observed in the radio channel as well as with the Auger SD433 array. The search is similar to a search performed with the prototype station located at the IceTop detector at the South Pole~\cite{IceCube:2021epf}. It consists of selecting radio waveforms with a high signal-to-noise ratio (SNR) from scintillator-triggered events and matching them with an SD event based on their trigger times and reconstructed directions. The matched events are then validated by resimulating the radio waveforms based on the air-shower reconstruction obtained from the SD. The search uses approximately three months of stable data taking: May 2023, August 2023, and January 2024. In May 2023, the DAQ operated at a sampling frequency of \SI{1}{GSps}. For the later months, the sampling frequency was lowered to \SI{800}{MSps}.

The search begins with the radio data, which is processed to remove various artifacts originating from the DAQ system. All traces, both scintillator- and software-triggered, are filtered to a \SI{110}{\mega\Hz} to \SI{185}{\mega\Hz} band. A further filter is applied to suppress RFI spikes in the background spectrum, as described in \refref{IceCube:2021qnf}. The SNR is calculated for each trace as the square of the ratio of the maximal signal amplitude over the RMS of the signal. The right panel of \reffig{fig:spectrum} shows the distribution of SNR values calculated for background waveforms. For each antenna and polarization, the 95th percentile value is calculated to define a threshold for the selection of high-SNR events.

To select candidates for radio waveforms containing an air-shower signal, scintillator-triggered events are required to have at least 1 polarization in each antenna having an SNR value larger than the 95\% value obtained from the background waveforms. A simple directional reconstruction is performed on all events that pass the selection by fitting a planar radio wavefront to the times of the signal peaks. Based on this, possible counterparts in the SD433 data can be found.

\begin{figure}
    \centering
    \includegraphics[width=0.5\textwidth]{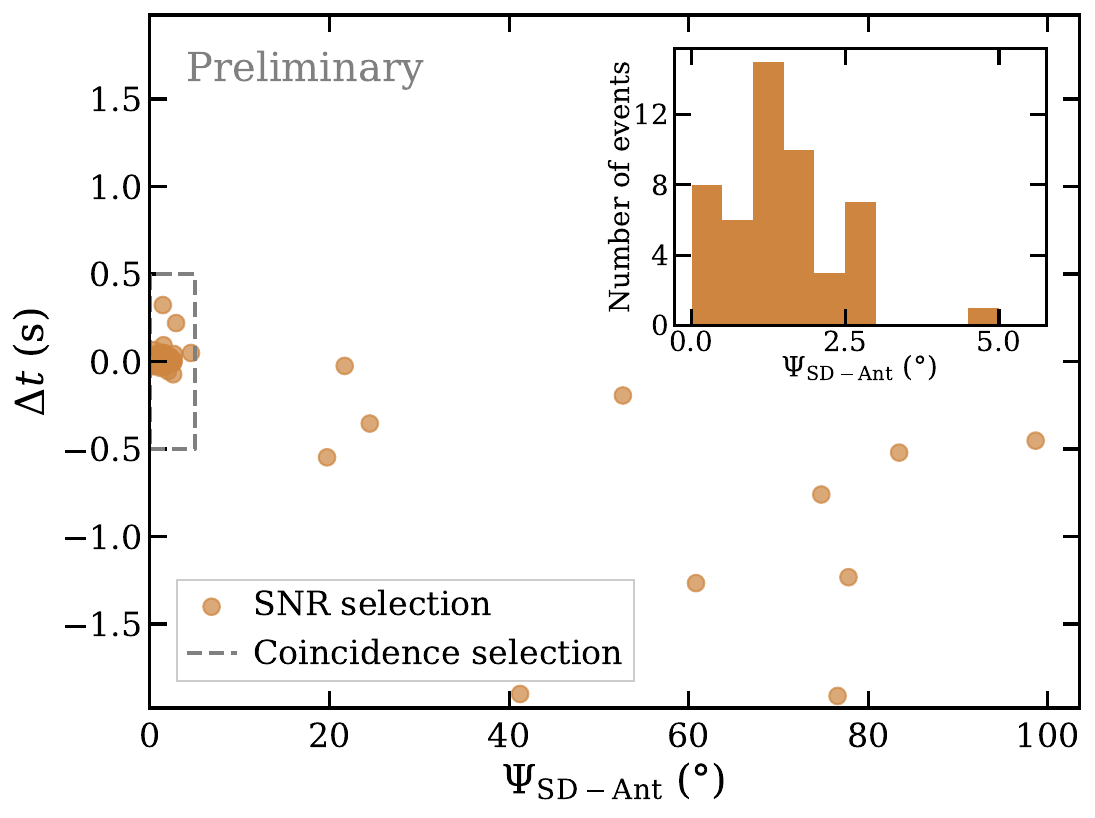}
    \caption{Scatter plot of the time offset and opening angle between air-shower candidates observed with the prototype station and events observed by Auger SD433. Events within the grey box are considered to be air showers observed by both detectors. The inset shows the distribution of opening angles between the directional reconstructions from SD433 and from the the radio signals.}
    \label{fig:coincidence}
\end{figure}

We use a preliminary SD433 dataset for the time period of the radio data, including preliminary directional and energy reconstructions. The energy estimation is described in \refref{PierreAuger:2023dju}. The global time offset between the prototype station DAQ and the SD433 array is determined on a statistical basis by comparing time differences between a large number of SD events and scintillator triggers. Due to a time synchronization issue in the prototype station, this is repeated on a daily basis. For each radio event candidate, SD events in a window of $\pm \SI{0.5}{\s}$ are selected. While this is a large window, the rate of events in the SD dataset is only of the order \SI{10}{\milli\hertz}. The directions reconstructed with the radio signals and the SD detector are then compared. If the difference is smaller than $5^\circ$, the events are considered to be caused by the same air shower. This coincident event selection is demonstrated for all radio event candidates in \reffig{fig:coincidence}. The distribution of opening angles between the reconstructions from the two detectors peaks around $1.5^\circ$.

\begin{figure}
    \centering
    \includegraphics[width=0.38\linewidth]{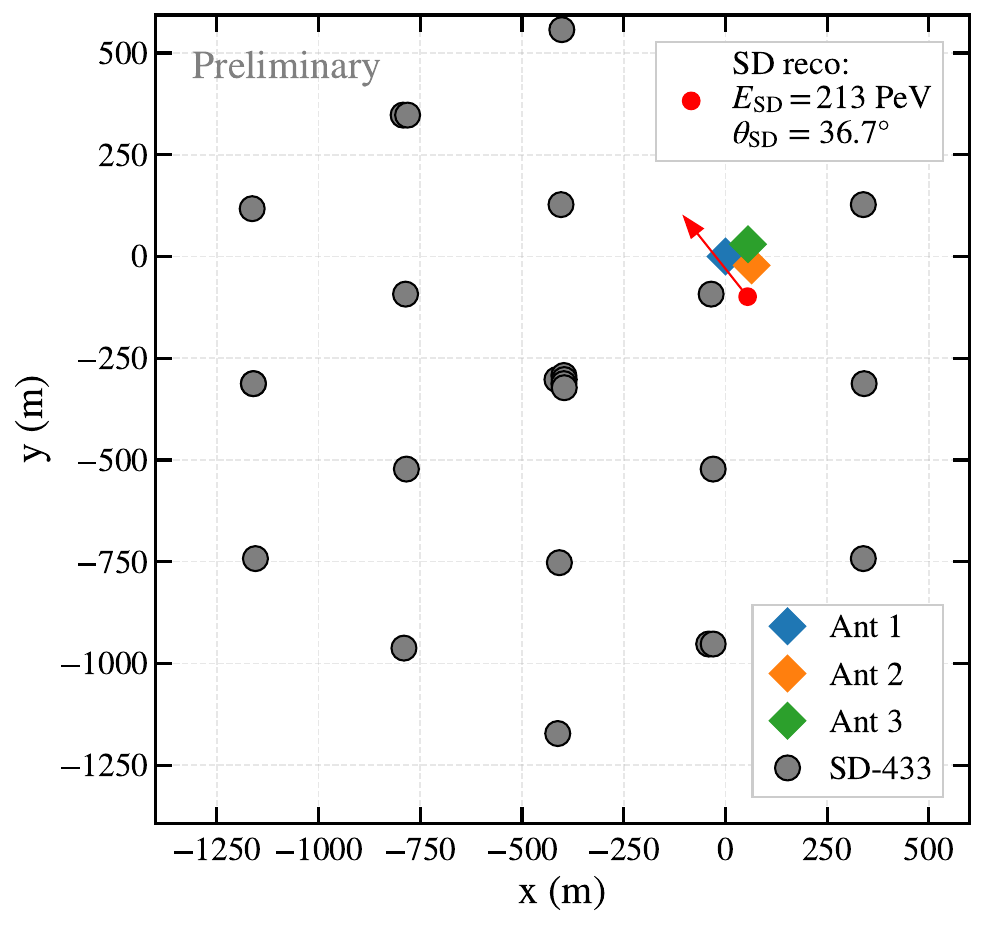}\hfill\includegraphics[width=0.55\linewidth]{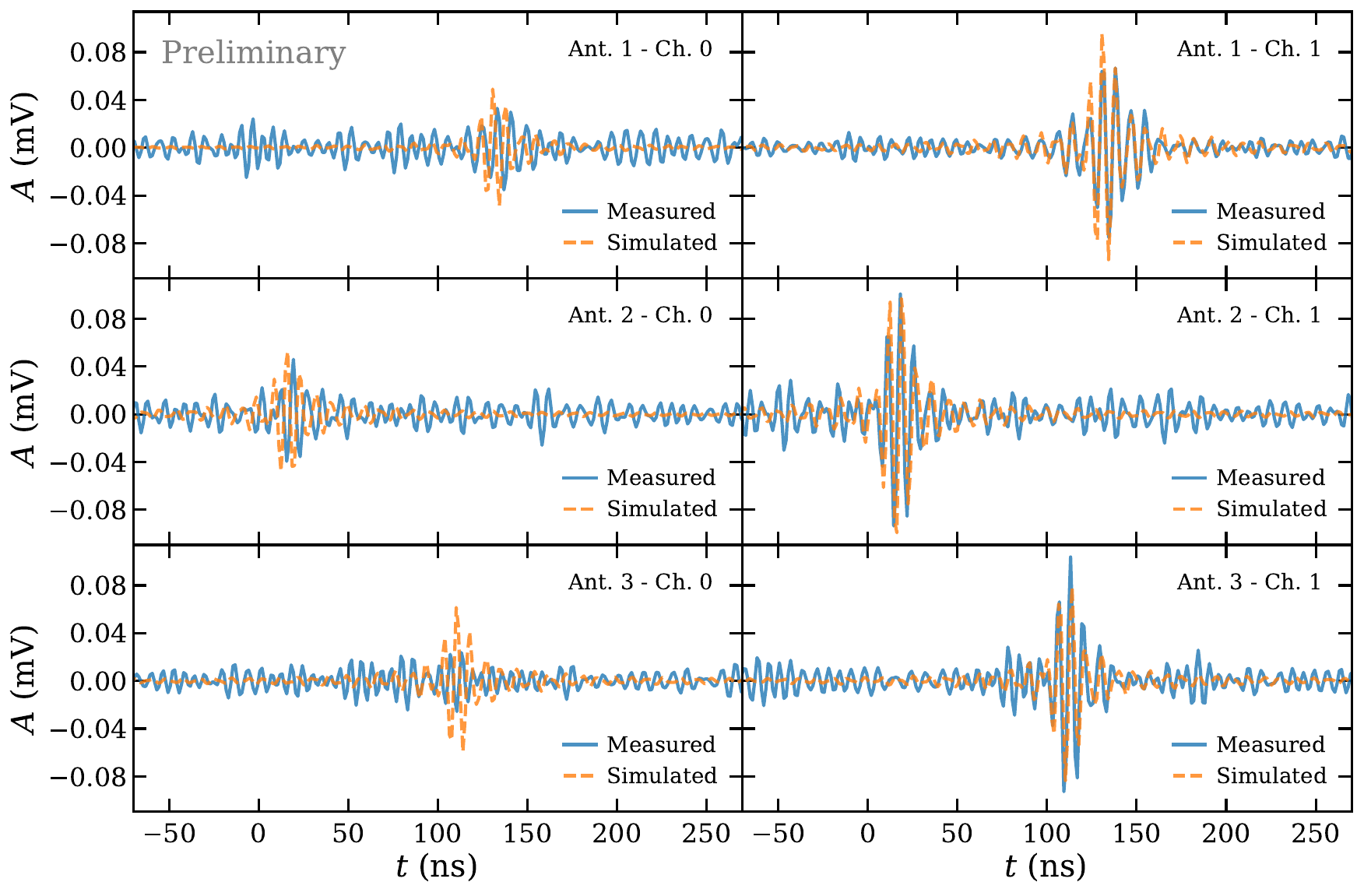}
    \caption{Example of an identified air shower. The left panel shows the reconstructed shower core position and direction as obtained from the SD433 array relative to the positions of the antennas and SD detectors. The right panel shows a comparison between the radio waveforms obtained from a simulation of the air shower, based on the SD reconstruction, and the traces recorded with the antennas.}
    \label{fig:example}
\end{figure}

To validate that the identified radio signals correspond to the air showers observed with SD433, the SD-reconstructions of the shower core position, direction, and energy are used to perform a simulation of the air-shower radio emission with CoREAS~\cite{Huege:2013vt}. The simulations are performed with an atmospheric model, geomagnetic field strength/direction, and observation level suitable for the Auger site. The response of the antenna to the simulated electric fields is then simulated, after which the electronics response is added to the signal. The resulting simulated waveform can then be compared to the observed one. An example event is shown in \reffig{fig:example}. The left panel shows the location of the radio antennas and the shower reconstruction in the SD433 array. The right panel shows a comparison of the simulated and observed waveforms, demonstrating a fairly good agreement between both. Note that uncertainties due to the reconstruction of the core position, shower direction, and energy as well as due to the depth of shower maximum are not yet taken into account and may further improve the level of agreement.

\begin{figure}[h]
    \centering
    \includegraphics[width=0.48\linewidth]{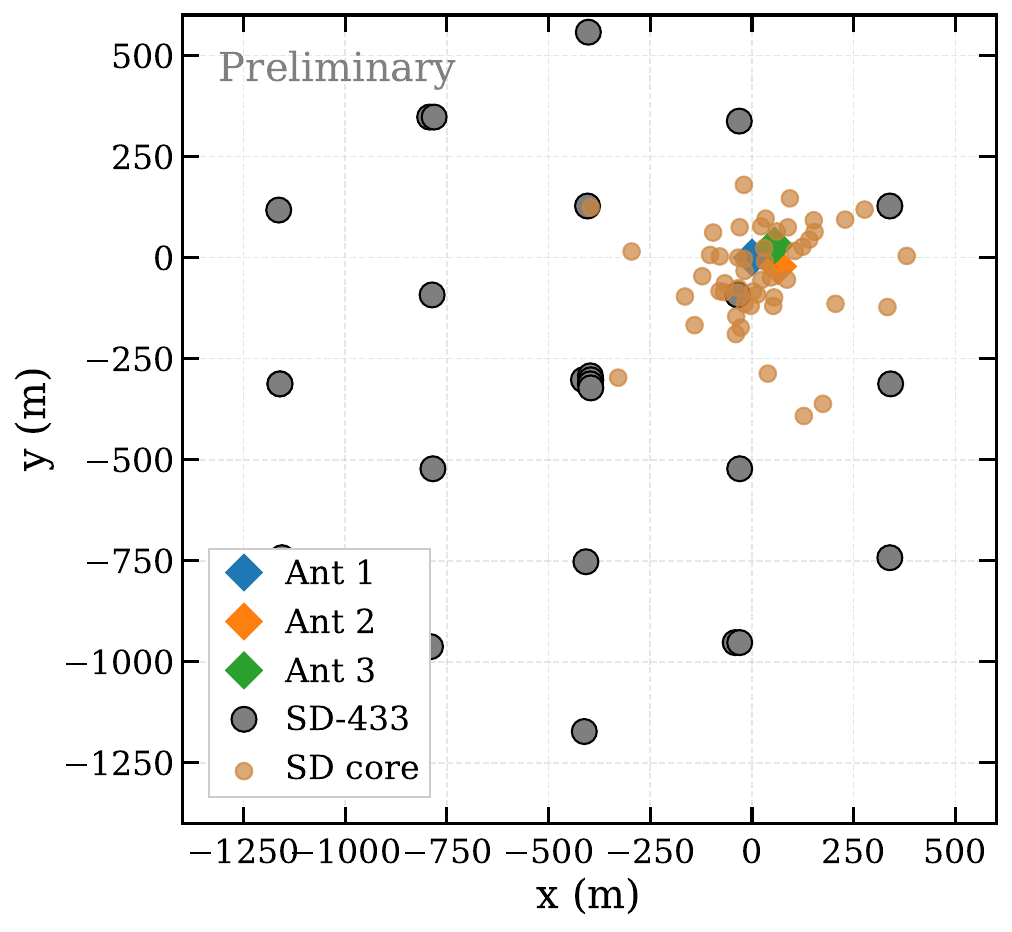}\hfill\includegraphics[width=0.48\linewidth]{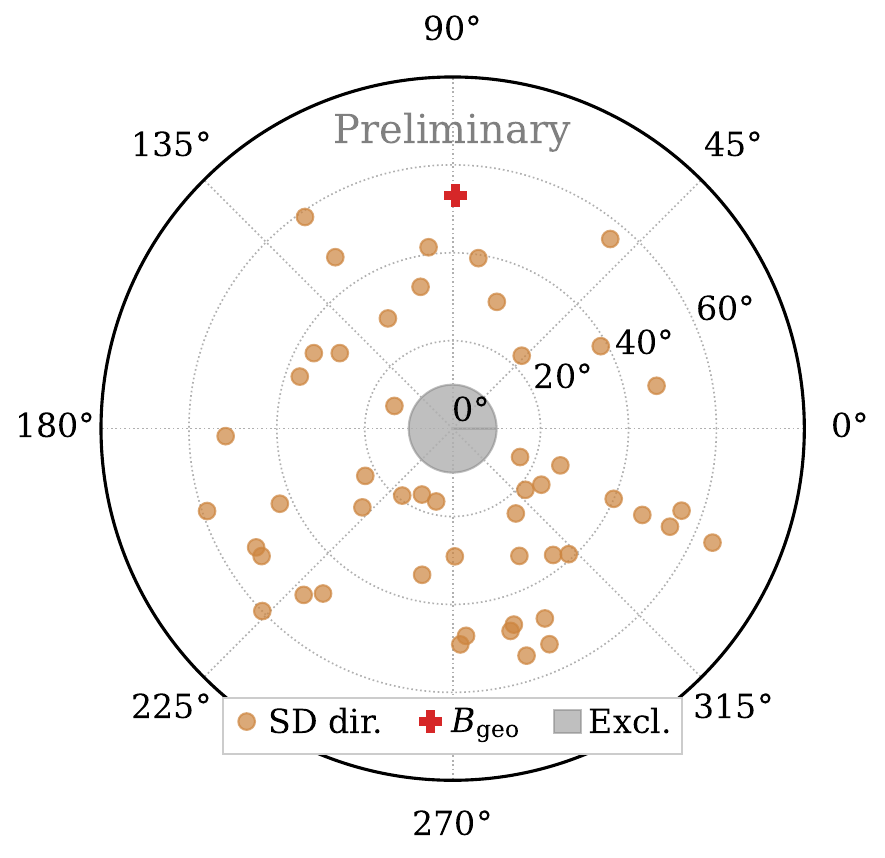}
    \caption{Left: Reconstructed shower core positions of the identified air-shower events. Right: Polar plot of the reconstructed directions of the identified events. The direction of the geomagnetic field is shown in red. The area of zenith angles smaller than $10^\circ$ was excluded from the analysis. Both plots show quantities reconstructed by SD433.}
    \label{fig:cores}
\end{figure}

\begin{figure}[h]
    \centering
    \includegraphics[width=0.5\textwidth]{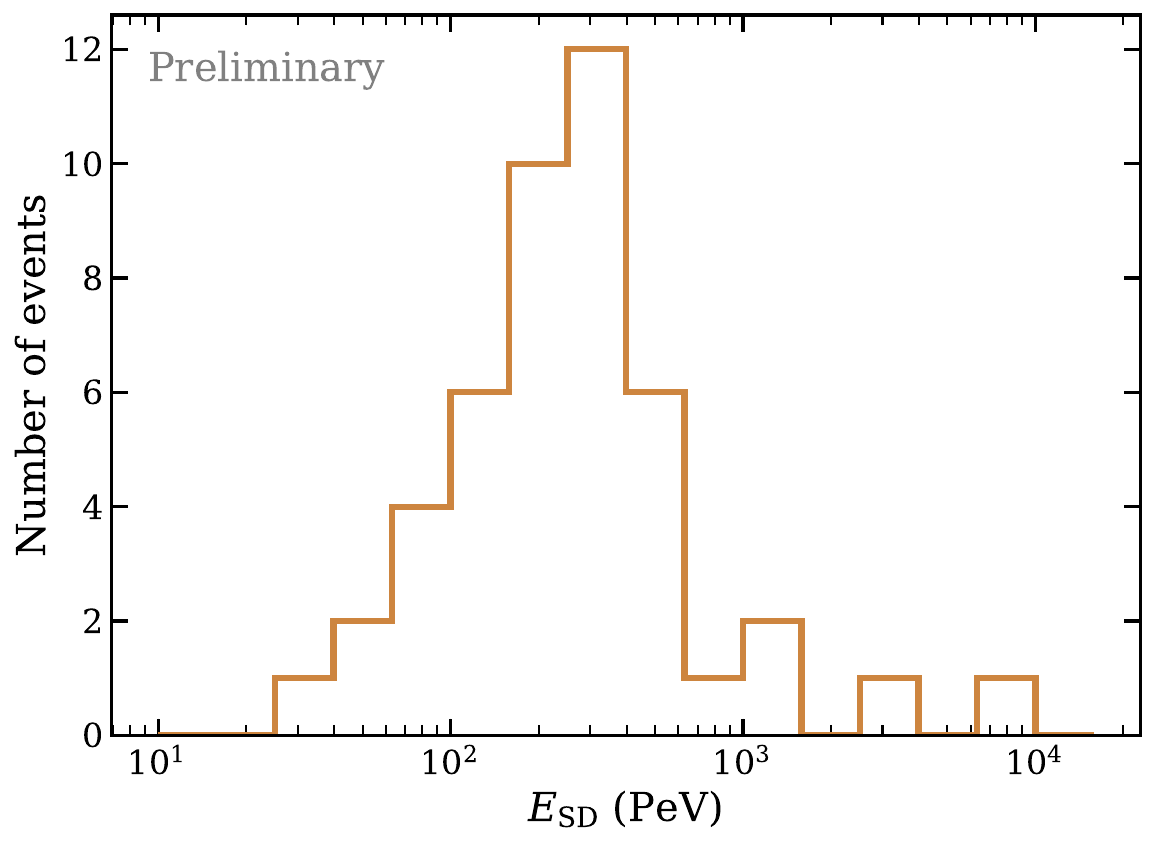}
    \caption{Distribution of the primary cosmic-ray energies estimated from the SD433 shower reconstruction for all identified air showers.}
    \label{fig:energy}
\end{figure}

A total of 50 events are identified in this way in approximately three months of data. \reffig{fig:cores} shows the distribution of the shower core positions and directions for these events as reconstructed by SD433. The core positions clearly cluster around the location of the prototype station. In the polar plot showing the reconstructed directions, a tendency for showers to be observed at larger angles with respect to the magnetic field direction is visible, as expected. The region of $10^\circ$ around the zenith is currently excluded from the analysis, as some persistent DAQ artifacts in the radio waveforms can mimic vertical air showers. The distribution of energies estimated from a preliminary SD433 reconstruction is shown in \reffig{fig:energy}. Showers are observed starting at several tens of PeV, and the distribution peaks around \SI{300}{\peta\eV}.

\section{Conclusion \& Outlook}

A prototype station of surface detectors for IceCube-Gen2 has been taking data at the site of the Pierre Auger Observatory since the end of 2022. In this contribution, we have presented the first results from the data recorded with the three radio antennas of the station. The data has been used to study the radio background at the Auger location in the band of \SI{70}{\mega\Hz} to \SI{350}{\mega\Hz}, confirming that we see Galactic noise in clean parts of the frequency band.
We have also demonstrated the first observation of air showers with this station, in coincidence with the Auger SD433 array. High-SNR radio signals in the \SI{110}{\mega\Hz} to \SI{185}{\mega\Hz} band were selected, and their times and reconstructed directions were compared to events reconstructed by the SD detector. A total of 50 events recorded in both detectors have been identified in about 3 months of data. Their estimated energies start at several tens of PeV, not yet including improvements expected by applying artificial neural networks for identifying and denoising radio pulses~\cite{Rehman}. 

The operation of a detector station at the Auger site provides opportunities for comparison with the current and future stations at the South Pole, e.g.~for cross calibration of air-shower measurements at both locations. The successful observation of air showers with the prototype has also motivated studying the possibility of a larger array of SKALA antennas at the Pierre Auger Observatory, which may enable the detection of vertical air showers with full efficiency, complementing the other Auger radio arrays which are mainly sensitive to inclined showers, i.e. AERA~\cite{PierreAuger:2018pmw} and the Auger Radio Detector~\cite{Huege:2023pfb}.

\bibliographystyle{ICRC}
\bibliography{main}

\end{document}